\documentclass{aa}  

\usepackage{graphicx}
\usepackage{txfonts}
\usepackage{graphics,graphicx}
\usepackage{amsmath}
\usepackage{amssymb}
\usepackage{tabularx}
\usepackage{graphics,graphicx}
\usepackage{txfonts}
\usepackage{longtable}
\usepackage{textcomp}
\usepackage{hhline}
\usepackage{arydshln}
\usepackage{multirow}
\usepackage{lscape}
\usepackage{array}
\usepackage{natbib}
\usepackage{setspace}
\usepackage{multicol}
\usepackage{pifont} 
\usepackage{wrapfig}
\usepackage{amsmath}
\usepackage{hyperref}
\usepackage{fancybox}
\usepackage{color}
\usepackage{subcaption}
\usepackage[dvipsnames]{xcolor}
\usepackage[export]{adjustbox}
\usepackage[normalem]{ulem}
\usepackage[separate-uncertainty=true,multi-part-units=single]{siunitx}
\usepackage{acronym}
\usepackage[capitalise]{cleveref}

\usepackage{titlesec}

\newacro{nir}[NIR]{near-infrared}
\newacro{mir}[MIR]{mid-infrared}
\newacro{fir}[FIR]{far-infrared}
\newacro{mm}[mm-range]{millimetre wavelength range}
\newacro{cm}[CM]{core-mantle}
\newacro{emt}[EMT]{effective medium theory}
\newacro{dda}[DDA]{discrete dipole approximation}
\newacro{mhd}[MHD]{magnetohydrodynamics}
\newacro{ism}[ISM]{interstellar medium}
\newacro{amr}[AMR]{adaptative mesh refinement}
\newacro{los}[LOS]{line of sight}
\newacro{rt}[RT]{radiative transfer}

\def\arcsec{\hbox{$^{\prime\prime}$}}

\def\d{\mathrm{d}}

\begin{document}

   \title{Pristine composition or size evolution: Can current dust models reproduce emissivities observed in nearby protostars?}
   \titlerunning{Composition or evolution: Can current dust models reproduce emissivities observed in nearby protostars}

   \subtitle{}

   \author{M.-A. Carpine\inst{1}\and A. Maury\inst{4,5,1} \and
           N. Ysard\inst{2,3} \and L. Cacciapuoti\inst{6} }

   \institute{ Université Paris-Saclay, Université Paris Cité, CEA, CNRS, AIM, 91191, Gif-sur-Yvette, France
         \\
         \email{marie-anne.carpine@cea.fr}
         \and IRAP, CNRS, Université de Toulouse, 9 avenue du Colonel Roche, 31028 Toulouse Cedex 4, France
         \and Universit{\'e} Paris-Saclay, CNRS, Institut d'Astrophysique Spatiale, 91405, Orsay, France
         \and
         Institute of Space Sciences (ICE), CSIC, Campus UAB, Carrer de Can Magrans s/n, E-08193 Barcelona, Spain
         \and ICREA, Pg. Lluís Companys 23, Barcelona, Spain
         \and European Southern Observatory, Alonso de Cordova 3107, Vitacura, Region Metropolitana de Santiago, Chile}
         
   \date{\today}

  \abstract
   {Interstellar medium dust is a crucial asset in many astronomical observations, as its emission permits the retrieval of the physical properties of astrophysical structures. Characterising dust grains present in the dense gas and in star-forming environments is also key to constrain the pristine conditions for planetary formation. However, dust properties, especially in young protostars, remain poorly characterised and are still debated: low dust emissivities observed in nearby protostars are, for example, not completely explained to this day.}
   {In this study, we aim to determine (i) whether it is possible to retrieve the dust properties from multi-wavelength observations of the dust emission towards embedded protostars, and (ii) the extent to which current dust models should be excluded from or can reproduce the observed values of the dust emissivity index in young embedded protostars.}
   {We perform radiative transfer computations of the thermal dust emission from a model protostellar envelope, considering different dust optical properties commonly used in the community. This allows us to explore the effects of dust composition on the spectral index, and compare these synthetic observations to literature data of dust emission in nearby protostars, to try and explain the variation in the emissivity index in observations.}
   {We find a large variation in the spectral index as the sole result of different dust models, without the need for dust grain size evolution. However, our work does not allow us to reproduce the lowest emissivity index values found in some protostellar envelopes without including unexpectedly large millimetre-sized processed grains. We show a tentative clustering of the emissivity index values depending on the star-forming region these protostars belong to, suggesting a potential relationship with the pristine dust reservoir at cloud scales.}
   {We show that appropriate methods, tested on theoretical models, allow us to measure the dust emissivity from observations of the spectral index at millimetre wavelengths with very little uncertainty, despite the large temperature and density gradients present in star-forming cores. Comparing the large magnitude of variation in emissivity index between the different observed sources with the dust models most commonly used by the community implies that the intrinsic composition of dust and its associated optical properties are not sufficient to explain the lowest spectral index values. Thus, early dust evolution producing larger dust grains in a dense medium may have to be taken into account to obtain a complete picture.}


   \maketitle


\section{Introduction}
\label{sec:intro}

Interstellar dust plays a crucial role in many astrophysical processes, yet understanding the nature and properties of dust grains remains a challenge. From the diffuse \ac{ism} to the dense medium of star-forming regions and all the way to planet-forming disks, sub-micronic dust grains \citep{mathis_size_1977,hirashita_synthesized_2013} undergo an evolution towards kilometre-sized planetesimals \citep{keppler_discovery_2018} while interacting with their environment. The surface of dust grains acts as a catalyst,for example, for the formation of complex molecules \citep{herbst_unusual_2021,dulieu_experimental_2010,Wakelam2017}. Dust grains are also key to heating and cooling processes; their interaction with the gas and coupling with the magnetic field \citep{zhao_formation_2020,maury_recent_2022} play a major role in angular momentum transport in star-forming regions, and thus for the formation of protostellar cores and rotating disks, and dust grains provide the building material for planetary formation. \\
Last but not least, most significantly here, dust provides a major observable in cold environments \citep{galliano_interstellar_2018}, and thus it is a great asset to probe the \ac{ism}, as well as dense star-forming regions. Dust grains interact with light through absorption, scattering, and emission, both in unpolarised and polarised fashion. In dense cold \ac{ism}, dust emission peaks at sub-millimetre and millimetre wavelengths \citep{planck_theory_1914,krugel_introduction_2008}, one of the ranges of interest for dust characterisation. Despite the dust being a good observable, the information one can retrieve from observations is scarce, and some questions still remain about dust properties and dust evolution throughout different environments. In the domain of dust thermal emission, a key parameter one can retrieve from observations is the 'emissivity index' $\beta$. This index is computed as the spectral slope of the absorption cross-section, which is also referred to as emissivity \citep{bohren_fundamentals_2006} and is approximated to a power law at large wavelengths when computing $\beta$. This index is known to be related to the size of grains \citep{testi_dust_2014}. In recent studies \citep{galametz_low_2019,cacciapuoti_faust_2025}, particularly low ($\beta<1$) emissivity indices have been observed in protostellar environments, possible evidence for grains with typical sizes \qty{100}{\um} \citep{ysard_grains_2019}, which are also needed to recover typical polarisation fractions and emissivity indices in  envelopes \citep{valdivia_indirect_2019}. \\
Beyond the low values, the emissivity index -- the spectral slope of the absorption cross-section -- displays a wide variety of values from source to source in these observations, with most values in the range from 0.4 to 1.7. Since differences in the emissivity index -- typically 1.5 to 2.5 -- are also found in the  \ac{ism} in different regions \citep{paradis_far-infrared_2011,paradis_modeling_2014,mason_confirmation_2020} which can be mostly reproduced by variations in the material composition, one can assume that the diversity of dust reservoirs in each different observed region is one of the causes of differences in the emissivity index. This illustrates that the dust emissivity index does not rely solely on grain size but on a variety of parameters \citep{kruegel_dust_1994}. Yet, in star-forming environments, variations in the dust emissivity index are almost always interpreted as a change in the dust grain size, whereas differences originating from the initial reservoir are usually not considered. Hence, between different regions of the \ac{ism} or different protostellar envelopes, the origin of dust and potential differences in the nature of grains make it difficult to conclude unequivocally whether different dust emissivity indices are due to different grain size and constrain the dust evolution. Dust evolution can be dispensable to explain some lower values of the emissivity indices, but we need to keep in mind that the lowest values of $\beta<1$ could only be explained by dust growth \citep{testi_dust_2014}.
Another key element in observational interpretation is the correct retrieval of the emissivity index. Protostellar envelopes constitute a dense, optically thick medium with temperature gradients, which strongly complicate the interpretation of dust emission and could make it a challenge to retrieve the dust properties.

Beyond observations, the interpretation of dust emission to derive physical quantities is based on the use of dust models. Such models help predict the thermal and polarised emission of dust by using expected dust properties and comparing them to observed results. For example, observational work often retrieves the dust mass from the dust thermal emission, using the following formula \citep{galliano_interstellar_2018,motte_f_circumstellar_2001}: $L_\nu(\lambda) = M_{\mathrm{dust}} \times \kappa_0 (\lambda_0/\lambda)^\beta \times  4\pi B_\nu(\lambda,T_{\mathrm{dust}} )$, assuming a single temperature on the line of sight, and with $\kappa_0 (\lambda_0/\lambda)^\beta$ the combined opacity of the dust population, intrinsically linked to dust properties and tabulated in dust models.
Several dust models have been developed over the years \citep{ossenkopf_dust_1994,weingartner_dust_2001,compiegne_global_2011,jones_evolution_2013,guillet_dust_2018,Siebenmorgen2023,hensley_astrodustpah_2023,ysard_themis_2024}. All these models can differ by the nature of the dust material with which they are built (silicate, carbon), the ratio of materials to each other, the stoichiometry (see the diversity of silicates in \citealt{ysard_themis_2024}), the crystalline structure of the materials or the structure of the grains (e.g. homogeneous, coated in a mantle, and so on). All these hypotheses about dust properties significantly change the optical properties, such as $\kappa$, used to derive dust masses from dust thermal emission, but also the way dust grains are heated and therefore the dust temperatures to be used to interpret the dust thermal emission.

Considering on the one hand the variety of emissivity indices observed in different protostellar sources, and on the other the variety of interstellar dust models with different optical properties existing in the current literature -- the most recent models exhibiting the lowest emissivity indices \citep{Siebenmorgen2023,hensley_astrodustpah_2023,ysard_themis_2024} -- this study is an effort to explore whether using different dust models to predict the dust thermal emission from embedded protostars at millimetre wavelengths may explain the observed diversity in emissivity indices, without dust grain evolution. Furthermore, we assess the ability to retrieve an intrinsic variation in dust properties, despite the large density and temperature gradients that affect the line of sight when measuring the spectral index in protostellar environments. Finally, we discuss expected changes in the dust emissivity spectral index as observed at millimetre wavelengths in protostellar envelopes, considering variations in the dust size distribution. Such predictions are particularly precious as they can be compared to the growing literature of measured dust emissivity indices in protostellar envelopes.

This paper is organised as follows: in \cref{sec:model}, we describe in detail the models and radiative transfer calculations performed. In \cref{sec:Results} we plot the results and derive emissivity indices of our computations for all the dust models considered. We then discuss those results in \cref{sec:Discussion}, and lastly we summarise them in \cref{sec:Conclusions}.

\section{Methods}
\label{sec:model}

The aim of this study is to investigate quantitatively how the choice of a given type of dust affects the resulting dust thermal emission and to explore the possible range of millimetre dust emissivities that could be realistically produced by state-of-the-art dust models used by the community. With this prospect, we chose to perform
synthetic sky models, combining a prototypical numerical \ac{mhd} model of a young embedded protostar and \ac{rt} computations of the dust thermal emission according to different dust models. 

\subsection{Template MHD model}
\label{subsec:MHD}

\begin{figure}[h]
  \centering
  \includegraphics[width=\linewidth,clip]{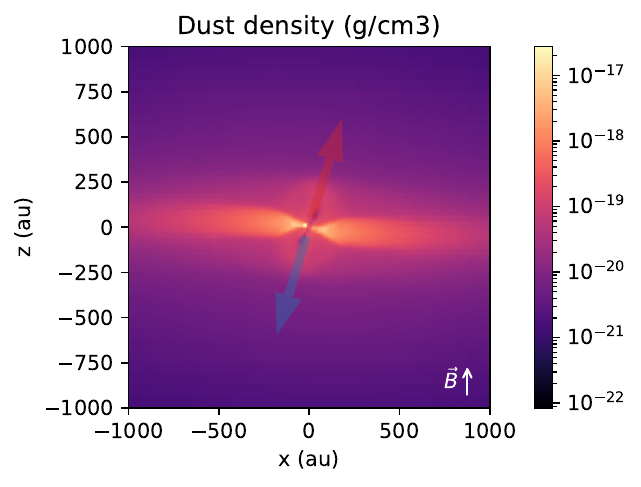}
  \caption{Dust density map of the template \ac{mhd} model used in this study, seen edge-on. In white, we represent the initial magnetic field direction ($z$-axis). The blue and red arrows represent the outflow direction, orthogonal to the midplane.
  }
  \label{fig:Dens}
\end{figure}

The template model used in this study is an output from the non-ideal \ac{mhd} simulation shown in \cite{hennebelle_what_2020}. 
It follows the evolution of the collapse of a protostellar core, using the RAMSES code \citet{teyssier_cosmological_2002, fromang_high_2006} and is described in great detail in \citet{valdivia_is_2022} (model 'ID R1' at $t_{0.1M_\odot}$). A brief description of the model is provided in \cref{sec:MHDdescr}. We used this model as a fiducial physical model for an isolated low-mass embedded protostar in the solar neighbourhood. From the 3D gas density distribution in the inner envelope, we plotted the corresponding dust density 2D map as observed in the edge-on configuration in \cref{fig:Dens}, assuming a dust-to-gas ratio 0.54 \% \citep{ysard_themis_2024}.

\begin{figure}[h]
  \centering
  \includegraphics[width=\linewidth,clip]{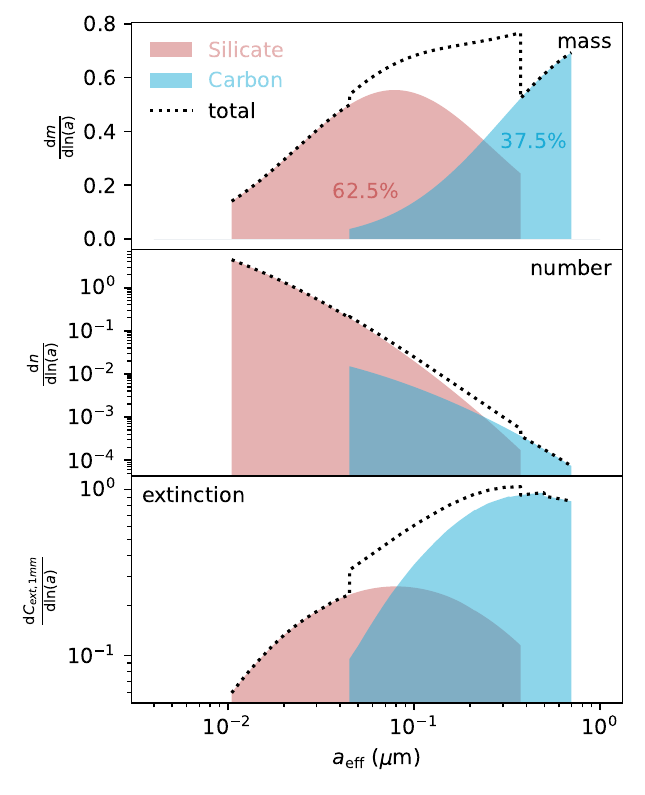}
  \caption{Dust size distributions in mass, number, and extinction cross-section $C_{\mathrm{ext}}$ at \qty{1}{\mm} (the latter is represented for the THEMIS 2.0 model as an illustration, but would be similar for the other models). The distributions are normalised, such that the integral of the total distribution equals 1 in each panel. 
  The size distribution is log-normal in number.}
  \label{fig:Size_dis_logn}
\end{figure}

\subsection{Dust model}
\label{subsec:dust}

\begin{table}[h]
\caption{Dust models used in this study.} 
\label{table:dustmodels}
\begin{tabular}{ccc}
\hline\hline
Model name    & Silicate grains    & Carbonaceous grains \\ \hline
              & $62.5 \%$          & $37.5 \%$           \\ 
              & 11 nm - 0.37$\mu$m & 45 nm - 0.7$\mu$m   \\ \hline\hline
Astrosilicate & Astrosilicate      & Graphite            \\ \hline
Astrodust     & Astrodust          & Astrodust           \\ \hline
THEMIS 2.0    & a-Sil + a-C mantle & a-CH + a-C mantle \\\hline
\end{tabular}
\tablefoot{Summary of the dust used for the three models used in this study, for the silicate and carbonaceous populations, respectively. We use the same mass proportion and size range for every model, which are specified in the second and third lines of the table.}
\end{table}

Our aim is to test the impact of the choice of different plausible materials for dust grains with varying optical properties on the resulting spectral index in the millimetre domain. We conducted our study using several dust models that have been shown to be successful in reproducing some of the observed dust emission properties in the diffuse \ac{ism}. We acknowledge that the typical dust grain in the diffuse \ac{ism} may not be the typical dust grain in dense protostellar environments, however, the \ac{ism} constitutes the starting point and the dust reservoir for protostellar envelopes, and many \ac{ism} dust models are conveniently available in the literature. Although the pristine dust grains populating the dense clouds and cores stem from the \ac{ism}, in these environments, they are probably covered by ice mantles and may be more porous if they started to stick to each other and grow. While we stress that the spectral emissivity index that we study in \cref{sec:Discussion} is only modified marginally and in a predictable way by such features \citep{Siebenmorgen2023, carpine_small_2025}, we propose here to investigate how good these pristine grains are at reproducing observed thermal dust emission in protostellar envelopes, and explore alternative routes that may help get closer to the observed dust spectral index for the current sample of protostars where it has been measured.
Therefore, we consider three widely used grain models: 
astrosilicates and graphites from \citet{weingartner_dust_2001} (Astrosilicate grains in the following), the astronomical dust model from \citet{hensley_astrodustpah_2023} (Astrodust grains in the following), and the THEMIS 2.0 dust grains as described in \citet{ysard_themis_2024} (THEMIS 2.0 in the following).
In the three models studied, the grains have different shapes and chemical compositions, leading to different optical properties. The relevant differences for our study are (i) the grain efficiency to absorb photons from the UV to the near-IR and (ii) the intrinsic spectral index of their extinction cross-section in the (sub-)millimetre. Given the same size distribution, these models produce grains with different equilibrium temperatures, which in turn lead to variations in both their thermal emission and the corresponding spectral index.

These models make different hypotheses regarding the relative distribution of carbons and silicates and use one or two populations of large grains.
\cite{weingartner_dust_2001} define two populations: one of graphite and one of the so-called astrosilicates \citep{draine_optical_1984}, whereas \cite{hensley_astrodustpah_2023} define only one population of 'astrodust', mainly a mixture of carbon and silicate. The model of \cite{ysard_themis_2024}, THEMIS 2.0, defines two populations, one of amorphous silicates (a-Sil)\footnote{In this study, we use the 'best fit' model detailed in Table 3 of this reference.} coated by a 5 nm-thick mantle of hydrogen-poor carbon (a-C) and one of amorphous hydrogen-rich carbon (a-CH) coated by a 20 nm-thick mantle of hydrogen-poor carbon. A summary of the dust populations is provided \cref{table:dustmodels}.

We use the same grain size distributions for all three models: each dust population (carbons and silicates) is represented by a log-normal size distribution identical to those presented in \citet{ysard_themis_2024}, with minimum-maximum grain sizes of 45 nm - \qty{.7}{\um} and 11 nm - \qty{0.374}{\um}, respectively. The dust size distribution is represented in \cref{fig:Size_dis_logn}. We note that ignoring the presence of a possible population of very small grains with sizes $\lesssim 10$ nm has no consequences on dust thermal emission at millimetre wavelengths in dense environments \citep[see for instance][Fig. 2, 3 and 9]{compiegne_global_2011}, therefore we do not include any very small grain populations. We also stress that according to \citet{ysard_grains_2019}, \citet{draine_submillimeter_2006}, and \citet{birnstiel_dust_2024}, a significant reduction of the dust emissivity index cannot be achieved without including very large grains $>\qty{100}{\um}$. The effect of increasing grain size is discussed in \cref{subsec:growth}. All models also use the same dust-to-gas mass ratio integrated over the full range of sizes of 0.54 \%.
For THEMIS 2.0, therefore, we take the model as it stands, removing the nanograin population and adjusting the mass proportions to 62.5 percent of silicate grains and 37.5 percent of carbon grains to match other models. For \citet{weingartner_dust_2001}, astrosilicates follow the same size distribution as the THEMIS 2.0 silicates, while the graphites follow the same size distribution as the THEMIS 2.0 carbons. For Astrodust, the two size distributions are used with the single population of grains in this model to achieve perfect equivalence with the two previous models. 
These size distributions are relatively far from the standard MRN \citep[Mathis, Rumpl, Nordsieck: ][]{mathis_size_1977}, but correlate better with the dust population that underwent collisional growth \citep{ysard_grains_2019, lorek_local_2018}. Overall, choosing between an MRN and a log-n grain size distribution, provided that they have similar boundaries, has little influence on the results presented in the following (\cref{sec:dustsize} shows a test carried out for comparison).

\subsection{Radiative transfer}
\label{subsec:RT}

\begin{figure}[t]
  \centering
  \includegraphics[width=\linewidth,clip]{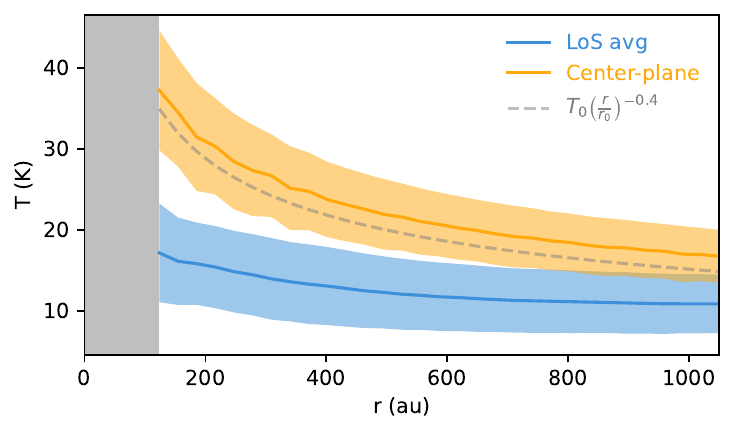}
  \caption{Dust temperature distribution obtained from the radiative transfer computation based on the THEMIS 2.0 dust model (note that other models provide similar dust temperature distributions). In yellow, temperature in the centre plane of the simulation ($y=0$), averaged in radius bins from the centre of the simulation, with the standard deviation in each bin as uncertainties. In blue, temperature averaged along the line of sight in each radius bin, weighted by the local dust density. The uncertainties are defined as the standard deviation of the temperature along each line of sight (weighted as well by the density) for each radius bin. In grey, the standard power-low model adopted in \citet{galametz_low_2019}. $T_0 = \qty{38}{\K}$, $r_0 = \qty{100}{au}$.}
  \label{fig:T}
\end{figure}

We populate our \ac{mhd} simulation with each of the three dust grain models and calculate the resulting dust temperature and thermal emission using the 3D radiative transfer code POLARIS \citep{reissl_radiative_2016}. 
We do not consider any photons from the outside (no interstellar radiation field) and assume that the heating source is the protostar itself, which we model with an equivalent young solar-type star with radius $1 R_\odot$, mass $1 M_\odot$, radiating like a blackbody at \qty{5804}{\K}. Moreover, to avoid issues of trapped photons in the very dense gas at scales <20 au (where the model also does not follow the physics accurately because of limited cell size in the AMR - 1 au at minimum), we artificially carve a \qty{4}{au} radius hole around the protostar where we set a density of zero. This standard practice allows photon propagation at reasonable computing costs in high-density environments \citep{valdivia_indirect_2019,le_gouellec_physical_2023,tung_accuracy_2024}. Typical size of cells is that from the \ac{mhd} model, ranging roughly from \qty{1}{au} to \qty{250}{au}, and having a typical (median) \qty{1}{mm} optical depth of \num{3e-5} per cell. Note that the whole plane of sky displays an optical depth of \qty{1}{mm} order of magnitude \num{1e-3} in the \qty{1000}{au} box, \num{1e-2} in the \qty{100}{au}, and $>0.1$ only $<\qty{10}{au}$ from the protostar.

The first step of the radiative transfer is the computation of the dust heating by Monte Carlo photon transfer. For this simulation, we used a total of 500000 emitted photons to ensure limited errors due to the Monte Carlo simulation (we find a temperature noise due to the Monte Carlo  standard deviation of $3\%$, which proves satisfactory convergence). The computed temperatures are represented in \cref{fig:T}\footnote{The temperature computed with THEMIS 2.0 model is given as an illustration here, but other models have temperatures that vary by less than 10\% from those results.}. We plotted both the temperatures in the centre plane and averaged along the line of sight, weighted by density
\begin{equation}
T(x,z) = \frac{\int T(x,y,z)\rho(x,y,z)\d y}{\int\rho(x,y,z)\d y},
\end{equation}
where $T$ is the dust temperature from the model and $\rho$ is the density of the gas mass. The integral covers $y$ from \qty{-8000}{au} to \qty{8000}{au}.
Overall, the temperatures in the cells range from \qty{80}{\K} (< \qty{20}{au} from the central protostar) to \qty{10}{\K} in the outer parts of the simulation. Note that most of the gas and dust grains located at >\qty{100}{au} from the central protostar have a temperature below \qty{20}{\K}.

In a second step, we used POLARIS to compute the intensity maps of the dust emission, using the ray-tracing part of the code. We chose to place the template protostar at a distance of \qty{250}{pc}, typical of the most-studied local star-forming regions. The total observed region is a \qty{8000}{au} square, sampled with \qty{5}{au} pixels.

\section{Results}
\label{sec:Results}

\begin{figure}[t]
  \centering
  \includegraphics[width=\linewidth,clip]{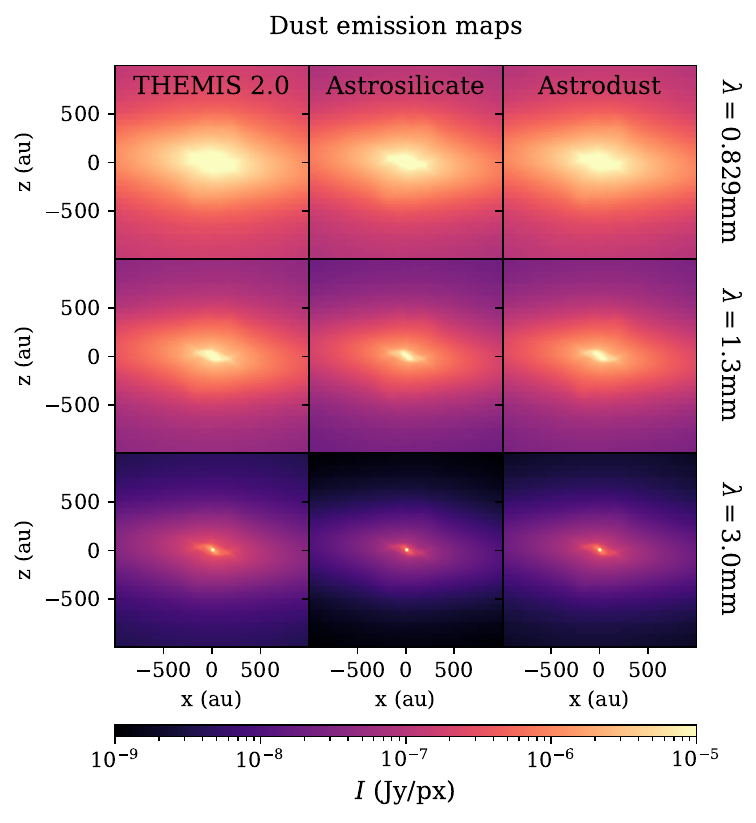}
  \caption{Flux density maps computed from the dust thermal emission from radiative transfer computation at different wavelengths (from top to bottom \qty{0.8}{\mm}, \qty{1.3}{\mm}, \qty{3}{\mm}). In each column, a different dust model is used (from left to right THEMIS 2.0, Astrosilicate, and Astrodust). Pixels are \qty{5}{au} in size.}
  \label{fig:I_0_Maps}
\end{figure}
\begin{figure}[h]
  \vspace{\floatsep}
  \centering
  \includegraphics[width=\linewidth,clip]{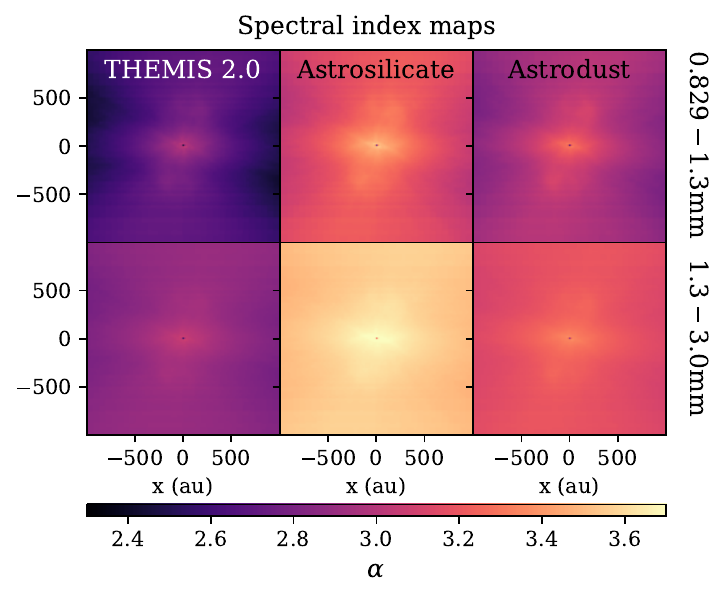}
  \caption{Emissivity index maps, derived from the intensity flux maps between different wavelength ranges (from top to bottom \qty{0.8}{\mm}-\qty{1.3}{\mm} and \qty{1.3}{\mm}-\qty{3}{\mm}). In each column, a different dust model is used (from left to right THEMIS 2.0, Astrosilicate, and Astrodust).}
  \label{fig:Alpha_Maps}
\end{figure}

In order to discuss the ability of the dust models to reproduce observed characteristics from protostellar envelopes, we have produced the sky models of dust thermal emission expected in key millimetre bands, widely used to probe protostellar environments. Synthetic dust continuum emission maps are shown in \cref{fig:I_0_Maps}. Each dust model (one per column) is used to produce the synthetic maps at three wavelengths: 0.8mm, 1.3mm, and 3.0mm (one per line). 
The synthetic sky maps show, as expected, dust thermal emission flux densities decreasing with longer wavelengths, as most of the dust present in these environments has low temperatures $T<50K$, and the millimetre wavelengths are located in the Rayleigh-Jeans regime for a modified greybody. We also note that the maps are qualitatively similar, with an increase of surface brightness by several orders of magnitude from the outer towards the centre of the envelope, due to the combined effects of the peaked density profile of the envelope, and of the increase in temperature due to heating from the central source. Yet we note quantitative differences in the emission from model to model and specifically in the spectral dependence: the intensity maps hint at a different decrease with wavelength for each model. To probe these differences, multi-wavelength maps are used to produce spectral indices maps ($\alpha$ maps) (plotted \cref{fig:Alpha_Maps} and later in \cref{fig:AlphaBeam,fig:BetavsAlpha}) as
\begin{align}
    \alpha &= \frac{\d \log I_\nu}{\d \log \nu}\\
     &\simeq  \frac{\log I_{\nu_1}-\log I_{\nu_2}}{\log \nu_1 - \log \nu_2} .
\end{align}

In the case of optically thin regions, if we assume a single dust temperature, the intensity can be rewritten
{\begin{align}
     I_\nu &= (1-\mathrm{e}^{-\tau_\nu})B_\nu(T)\\
     &\simeq  \tau_\nu B_\nu(T) \\
     &\propto (\nu/\nu_0)^\beta B_\nu(T).
\end{align}

Here we introduce the emissivity index $\beta$, following the commonly used hypothesis that the dust emissivity can be approximated by a power law with frequency at long wavelengths. This power-law model is a simplistic assumption which does not hold perfectly: for instance, the true dust emissivity can have different slopes in different wavelength regimes or even depend on the dust temperature (see \citealt{meny_far-infrared_2007,paradis_far-infrared_2011,demyk_low_2017}). Such effects are briefly discussed in \cref{sec:slopebreaks}. Because the objects studied (protostellar cores) are mostly populated by cold dust (5-20K) and also contain temperature gradients, the emissivity index $\beta$ cannot be derived directly from the observable $\alpha$. A temperature correction is required to disentangle the dust properties from the temperature effects (see \cref{sec:temperature} and \cref{eq:beta}). We test here whether the typical corrections applied to observations (see, e.g.  \citealt{galametz_low_2019,cacciapuoti_faust_2025,bracco_probing_2017}), where the 3D structure of the source is not known, hold in the specific case of protostellar envelopes and allow us to correctly retrieve the dust emissivity, using our models.

Thus, with the aim of investigating the robustness of measurements of the sub-millimetre emissivity index in protostellar envelopes, we transform the spectral index maps into dust emissivity index maps ($\beta$ maps) by correcting the spectral index from the temperature dependence, using a Planck law representative of the line-of-sight temperature in the model, as follows

\begin{align}
    \beta &= \alpha - \frac{\d \log B_\nu}{\d \log \nu} \\
    &= \alpha - \frac{\log B_{\nu_1}-\log B_{\nu_2}}{\log \nu_1 - \log \nu_2}, \label{eq:beta}
\end{align}
where $B_{\nu}$ is Planck's law (spectral radiance) computed by a density-weighted average along the line of sight ($y$-axis) 
\begin{equation}
B_{\nu}(x,z) = \frac{\int B_{\nu}(T(x,y,z))\rho(x,y,z)\d y}{\int\rho(x,y,z)\d y},
\end{equation}

where $T$ is the dust temperature from the model, and $\rho$ is the gas mass density. The integral covers $y$ from \qty{-8000}{au} to \qty{8000}{au}.

We note that the dust emissivity index is sometimes computed with the simpler formula $\beta_\mathrm{R-J} = \alpha - 2$ \citep{nozari_peculiar_2025, kwon_circumstellar_2009}, which is only valid in the Rayleigh-Jeans approximation. In our case, dust temperatures are too low for the use of this approximation: in such a case, observers often use a 1D radial variation for the temperature as an approximation to correct from $\alpha$ to $\beta$ \citep{galametz_low_2019,cacciapuoti_faust_2025}. Here we take advantage of knowing the full 3D structure in our models to test whether a full computation of the mean temperature in 3D, integrated along the line of sight, makes a difference in the measured $\beta$ values.
More details and a comparison of the values are provided in \cref{sec:temperature}, and are discussed extensively in \cref{sec:tempunderest}. The emissivity index values retrieved from the simulation are later plotted and discussed in \cref{sec:Discussion}. 

The maps of the spectral index recovered from the radiative transfer using the different dust models are shown in \cref{fig:Alpha_Maps}. The upper panel shows the spectral index computed between 0.8~mm and 1.3~mm, while the lower panels show the spectral indices computed between 1.3~mm and 3~mm for each of the three dust models explored in this study. 
\Cref{fig:Alpha_Maps} shows the diversity of spectral indices produced by the different dust models, looking at a single object, with the Astrosilicate model showing the largest values, up to $\alpha$ $\sim 3.6$, and the THEMIS 2.0 model showing the flattest spectral dependence, with $\alpha$ values as low as $\sim 2.4$.

\section{Discussion}
\label{sec:Discussion}

\begin{figure}[t]
  \centering
  \includegraphics[width=\linewidth,clip]{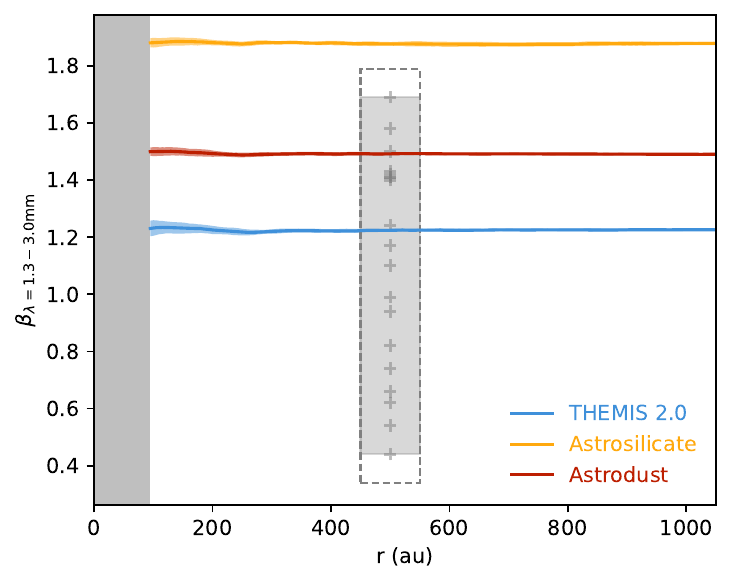}
  \caption{Comparison of the radial emissivity index for different dust optical properties. In colours, data from the radiative transfer computations run in this work. In grey, data retrieved from observations in \citet{galametz_low_2019} and \citet{cacciapuoti_faust_2025}. In observations, the emissivity index is given at \qty{500}{au} scale for reference value (plus marker). In dashed grey, we represent the 0.1 uncertainty due to 1D temperature modelling in observations.}
  \label{fig:Beta_modVSobs}
\end{figure}

The spectral indices derived in \cref{sec:Results} display large variations. In this section, we discuss possible reasons for the differences in the derived emissivity index relative to the dust optical properties, their variation with wavelength, and how they compare to values derived from observations of embedded protostars.

In \cref{sec:Results}, we plotted the observed spectral index that we then transform into emissivity indices, with the process described in \cref{eq:beta} to compare with the values from observations. Thus, even if the emissivity index is computed from an approximation assuming a constant slope beta at millimetre wavelengths, we can disentangle the bare properties of dust from any unwanted temperature effect on the observables, as discussed in \cref{sec:temperature}. For visualisation purposes, we present in \cref{fig:Beta_modVSobs} the radial profiles of the emissivity index, averaged in concentric bins centred on the source in the plane of sky. From here on, references to 'radial' dependency assume a dependency with surface radius in the plane of sky. For comparison, we also show observational results from \citet{galametz_low_2019} and \citet{cacciapuoti_faust_2025}, both of which derive a $\beta$ with the same formula as in \cref{eq:beta} and a 1D temperature approximation described hereinafter in \cref{sec:tempunderest}. These studies derived the emissivity index in the 1.3–3.2 mm and 1.2–3.1 mm wavelength ranges, respectively, for Class 0/I protostellar sources observed as part of the CALYPSO survey and the ALMA FAUST Large Program. Note that from \citet{cacciapuoti_faust_2025}, we excluded the sources IRAS4C and RCrA-IRS7B because the 3 millimetre emission from the envelope was not detected in the visibility datasets.
We stress that in our models the effects of wavelength-dependent interferometric filtering have not been taken into account, yet they should not affect the results as \citet{galametz_low_2019} and \citet{cacciapuoti_faust_2025} derive the emissivity indices from the visibilities, thus not being sensitive to these effects of flux filtering that would otherwise typically affect the dust emission maps at scales $\gtrsim$ 2000 au.

\subsection{Temperature estimation and emissivity index}
\label{sec:tempunderest}

First, we want to stress here that the values retrieved in \cref{fig:Beta_modVSobs} correspond, for each dust model, to the intrinsic value of the dust model’s emissivity index (computed from the integrated dust cross-section at millimetre wavelengths, tabulated in each respective model) with a precision of $1\%$. This shows that, for small grains in protostellar environments, observations of thermal dust emission at multiple millimetre wavelengths can be robustly used to measure the dust emissivity index if following the methodology described in \cref{sec:Results} to properly take into account temperature effects.

We stress that observational works (see, e.g. \citealt{galametz_low_2019,cacciapuoti_faust_2025}), correct the observed $\alpha_{mm}$ (projected in the plane of sky) thanks to an analytical 1D temperature profile estimated from the source luminosity, as the 3D information cannot be retrieved

\begin{equation}
    \label{eq:tempest}
    T(r) = 38 \left( \frac{L}{L_\odot}\right)^{0.2}\left( \frac{r}{\qty{100}{au}}\right)^{-0.4} K.
\end{equation}
This model is a good approximation of the temperature in three dimensions (as one can see in \cref{fig:T}), but does not quite capture the effects of temperature variation and `flattening' along the line of sight (see \cref{fig:T}). This leads to an overestimation of the temperature, which could in turn lead to an underestimation of the real emissivity index (see, for instance, the difference between $\beta_{T\mathbf{los}}$ and $\beta_{T\mathbf{cen}}$ in \cref{fig:BetavsAlpha}, which is about 0.1). 
However, if anything, the methods used from observations to probe the dust emissivity lead to biases only slightly underestimating $\beta$. If we were to re-estimate the $\beta$ from observations with a 3D averaging of the Planck's function along the line of sight, using \cref{eq:tempest}, instead of the 1D computation in the original studies, we find that $\beta$ values would only vary by $[0.02 - 0.07]$. Our analysis thus allows us to confirm, for example, that in the case of the analysis carried out in \cite{galametz_low_2019,cacciapuoti_faust_2025}, the $\beta$ values obtained using the 1D temperature correction allow us to trace a true distribution of the emissivity indices $\beta$, within 0.1, and that the range of values obtained is unlikely to be due, for example, to different biases linked with different temperature profiles. 
Beyond these specific studies, our work confirms that the emissivity index $\beta$ of the dust in protostellar envelopes can be retrieved from observations of the spectral index $\alpha_{mm}$, at all scales 100-1000 au, at least for solar-type protostars. Observed variations of the spectral index $\beta_{mm}$ in these environments can be linked to variations in intrinsic dust properties, as long as effects due to optical depth (removal of an optically thick compact component) and temperature (correction using an averaged Planck law) are properly taken into account, even with simple assumptions.

\subsection{Observed gradient of the spectral index}

Our radiative transfer computations suggest, as seen in \Cref{fig:Beta_modVSobs}, that a steady dust emissivity index should be observed everywhere in the envelope, from radii 200 au up to 1000 au.
However, many of the sources (13 out of 19) observed in \citet{galametz_low_2019,cacciapuoti_faust_2025} exhibit a gradient of $\beta$ that is mostly lower than $\beta$ towards the centre of the envelope. 

We stress that our models include a single dust population identical in size distribution throughout the protostellar envelope, leading to the homogeneous dust emissivity index visible in \cref{fig:Beta_modVSobs}. Observed protostars, on the other hand, are likely to host dust grains with locally varying sizes due to early dust evolution up to sub-millimetre size. Since the typical timescale for dust evolution shrinks at a higher gas density, larger grains could be expected in the inner envelope, showcasing lower $\beta$. The approximation on the model, which currently does not account for the complexity of dust spatial variations,  leads to discrepancies in the gradient of emissivity indices.

We also note that, if the modelled emissivity index is consistent throughout the grid, the spectral index $\alpha$ shows radial structure, due to the temperature effects affecting the relationship between $\alpha$ and $\beta$. This is discussed in more detail in \cref{sec:temperature}. 

\subsection{Wavelength dependency of the spectral index}
\label{sec:slopebreaks}

We emphasise that the emissivity index characterises the slope of the dust emissivity at large wavelengths, where it is typically assumed to behave as a decreasing power law. This is an approximation, and one cannot directly measure the dust emissivity, which does not strictly behave as a spectral power law. However, this approximation is required in observations in which the spectral resolution is limited and few bands in a large range permit the derivation of this index.
Thus, we stress that emissivities are not immune to slope breaks \citep[e.g.][]{boudet_temperature_2005,coupeaud_low-temperature_2011,demyk_low-temperature_2017,demyk_low_2017,carpine_small_2025}, and that the wavelength range of computation of the emissivity index $\beta$ from \cref{eq:beta} is a determining factor and could lead to different results. \Cref{fig:BetavsAlpha} provides a good example of this variability (compare the solid lines in the left versus right panel, for each model), as it shows $\beta$ computed between 0.8-1.3~mm and 1.3-3~mm. In between the two ranges, all three models show a variation in $\beta$, with the Astrosilicate model having an increase in $\beta$ with wavelength (from 1.80 to 1.87), while the Astrodust and THEMIS 2.0 show a decrease (1.56 to 1.49 and 1.32 to 1.22, respectively). These slope breaks have several potential causes, from the optical properties of the grains to the effects of mixing several dust populations with different grain sizes.
This demonstrates the need to remain careful when comparing emissivity indices, whether they are from models or observations, and to keep comparisons to the same ranges of wavelengths.

\subsection{Variations of dust emissivity due to dust optical properties}
\label{subsec:dust_on_beta}

Besides dependencies on the radius and the wavelength, the emissivity index (\cref{fig:Beta_modVSobs}) displays a large offset between models and between observations.
In this section, we examine in greater detail the effects of varying dust properties on the `absolute value' of the emissivity index. As a reminder, all radiative transfer computations are performed on the same density grid and use an identical dust size distribution. The only parameter that differs between models is the set of optical efficiencies provided as input, that is, the intrinsic optical properties related to the material composition and shape of the dust grains. 

Radiative transfer computations reveal significant variations in the emissivity index $\beta$ between different dust models (\cref{fig:Beta_modVSobs}). For example, $\beta_{\lambda = 1.3\text{–}3,\mathrm{mm}}$ differs by up to 35\% (0.65) between THEMIS 2.0 and Astrosilicate models, with values ranging from approximately 1.225 to 1.877.
Since all models share the same grain size distribution, this diversity in $\beta$ can only be attributed to the intrinsic properties of the grains: namely, their composition and shape. These differences arise from both the nature of the constituent materials and the proportions in which they are combined.
In the Astrosilicate model, carbonaceous and silicate grains are treated as distinct populations. In contrast, the THEMIS 2.0 model features silicate grains with a carbonaceous mantle, whereas the Astrodust model mixes all components uniformly, without distinguishing between carbon-based and silicate-based grains. Even subtle variations in material properties can impact the emissivity index. As detailed in \cref{sec:asil}, the THEMIS 2.0 a-Sil7 and a-Sil2 models, which differ only in silicate stoichiometry, yield different values of $\beta_{\lambda = 0.8\text{–}1.3,\mathrm{mm}}$, around 1.23 and 1.32, respectively.

In addition to the diversity of emissivity indices predicted by the models, the observational values shown in \cref{fig:Beta_modVSobs} also display significant variation. These observations, obtained from sources located in different star-forming regions, likely reflect a range of dust reservoirs with distinct initial compositions or different dust evolution. Each of these environments could potentially be represented by different dust optical property models. It should also be noted that two of the most recent models, THEMIS 2.0 and Astrodust, both calibrated against Planck data, have much lower emissivity indices than Astrosilicate: these models provide a possible explanation for the values of $\beta \sim 1.2-1.6$ observed towards protostars (\cref{fig:Beta_modVSobs}).

\subsection{Reaching for the lowest: Indications of grain size evolution?}
\label{subsec:growth}
 
This study shows that a wide range of $\beta$ values can be achieved solely by modifying the dust composition. In particular, we demonstrate that $\beta$  at millimetre wavelength can be reduced by up to 35\% from the largest value in the considered models, reaching values as low as 1.25 only including small, sub-micron-sized grains.
However, some observational studies, such as those by \citet{galametz_low_2019} and \citet{cacciapuoti_faust_2025}, report even lower values, with $\beta < 1$ in several (8 out of 20) protostellar sources. Such values are not reproduced by our current synthetic observations using state-of-the-art small-grain-size dust optical properties, and cannot be explained by the temperature effect described in \cref{sec:tempunderest}. Neither does an underestimation of 0.1 of the observed $\beta$ due to the temperature estimation discussed in \cref{sec:tempunderest} explain the several very low $\beta$ values observed.
This discrepancy suggests that dust evolution may play a role, potentially involving grain growth into aggregates bigger than \qty{1}{\um}, to further significantly reduce the emissivity index. Indeed, at wavelengths of millimetre order, which is much larger than the grain size currently considered, decreasing the emissivity index to <1 values could require larger grains, up to a millimetre in size \citep{ysard_grains_2019, testi_dust_2014, birnstiel_dust_2024}.

\begin{figure}[t]
  \centering
  \includegraphics[width=\linewidth,clip]{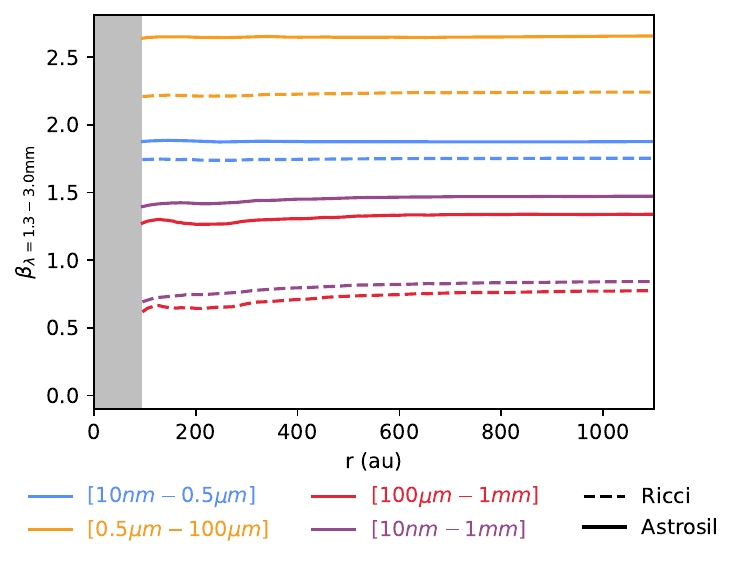}
  \caption{Comparison of the radial emissivity index for different dust optical properties from disk-like grains, unrealistic in the case of protostellar envelopes, with different size limits for the Astrosilicate (yellow) and Ricci (purple) dust, considering a power-law distribution ($q=3.5$). Size ranges in the legend are indicative (order of magnitude) and not exact.}
  \label{fig:Beta_modVSobsSIZE}
\end{figure}

To evaluate the possible decrease of $\beta$ with dust size, we investigated a variation in the minimum and maximal dust grain sizes. Among the three previously used \ac{ism} models, only Astrosilicate provides optical properties of grains up to millimetre sizes. We repeated the same radiative transfer computations for those grains, this time varying both minimum and maximum grain sizes to 'medium' (\qty{0.37}{\um}-\qty{100}{\um} for silicates, \qty{0.7}{\um}-\qty{100}{\um} for carbonaceous, all labelled $[\qty{0.5}{\um}-\qty{100}{\um}]$), 'large' (\qty{100}{\um}-\qty{1}{mm} for all grains), and 'all sizes' (\qty{0.37}{\um}-\qty{1}{mm} for silicates, \qty{0.7}{\um}-\qty{1}{mm} for carbonaceous, all labelled $[\qty{0.5}{\um}-\qty{1}{mm}]$) in addition to the "small" (as described in \cref{subsec:dust}, and labelled $[\qty{10}{nm} - \qty{0.5}{\um}]$) grain sizes. For completeness, we also conducted the study with an additional model of processed dust based on Astrosilicate from \citet{ricci_dust_2010} (Ricci in the following), widely used by the disk community, and which includes both porosity and ice elements. In every case for those new computations, we used a power-law size distribution with exponent $q=-3.5$, to match what has been done previously in the literature, and avoid any complex shifting of the log-normal distribution. We plot the resulting radial emissivity indices in \cref{fig:Beta_modVSobsSIZE}. We make the following observations, in line with what is suggested in the literature: increasing dust size with those models to \qty{100}{\um} is not sufficient to lower the emissivity index at the observed level, and we even increase the $\beta$ at those sizes compared with the 'small' population. When increasing up to millimetre 'large' grains, we do notice a significant decrease in the emissivity index, which is moderated if smaller grains are added ('all' population). Specifically, we reach low values of $\beta<1$ only with a population that includes grains of \qty{1}{mm} size with Ricci grains. This conclusion supports that of \citet{kruegel_dust_1994}, in which the authors found significant flattening of the dust opacity in the sub-millimetre and millimetre range when considering millimetre-sized grains. Note that the central decrease of the index is likely due to uncorrected optical depth effects, as the environment becomes very optically thick with such large grains. 

In any case, our study shows that, in the conditions reigning in protostellar envelopes, only including very large dust grains allows current models to reach the observed emissivities and that these cannot be due to biases in the optical depth or temperature variation. However, such a large fraction of large millimetre-sized grains is not a satisfactory solution to reproduce the observed emissivities, as the required level of grain growth is far beyond what is currently expected in protostellar envelopes: state-of-the-art models of grain growth are not able to grow grains to millimetre sizes at the densities $<10^6$ cm$^{-3}$ which prevail in the cores, in less than \qty{1}{\mega yr} \citep{ormel_dust_2009,silsbee_dust_2022,lombart_3d_2026}. We are therefore faced with the observation that the observed low emissivity indices in protostellar envelopes cannot currently be explained using expected dust grain sizes from current growth models. Modern models of dust aggregates, taking into account more complexity for the dust structure, and new optical properties (e.g. Astrodust or THEMIS 2.0), will undoubtedly help lift this current inconsistency. However, calculating the optical properties of large aggregates is beyond the scope of the current study.

It is also interesting to note that, depending on the choice of starting point for the composition of the grains (i.e. their optical properties), the degree of growth required to produce a similarly low beta may not be the same in all protostars, as the emissivity of the pristine monomers composing the larger grain, for example due to composition, may vary from region to region of the Galaxy.

\subsection{Investigating physical processes driving the dust emissivity variations observed in nearby protostars.}
\label{subsec:inhomogeneity}

\begin{figure*}[t]
  \centering
  \includegraphics[width=\linewidth,clip]{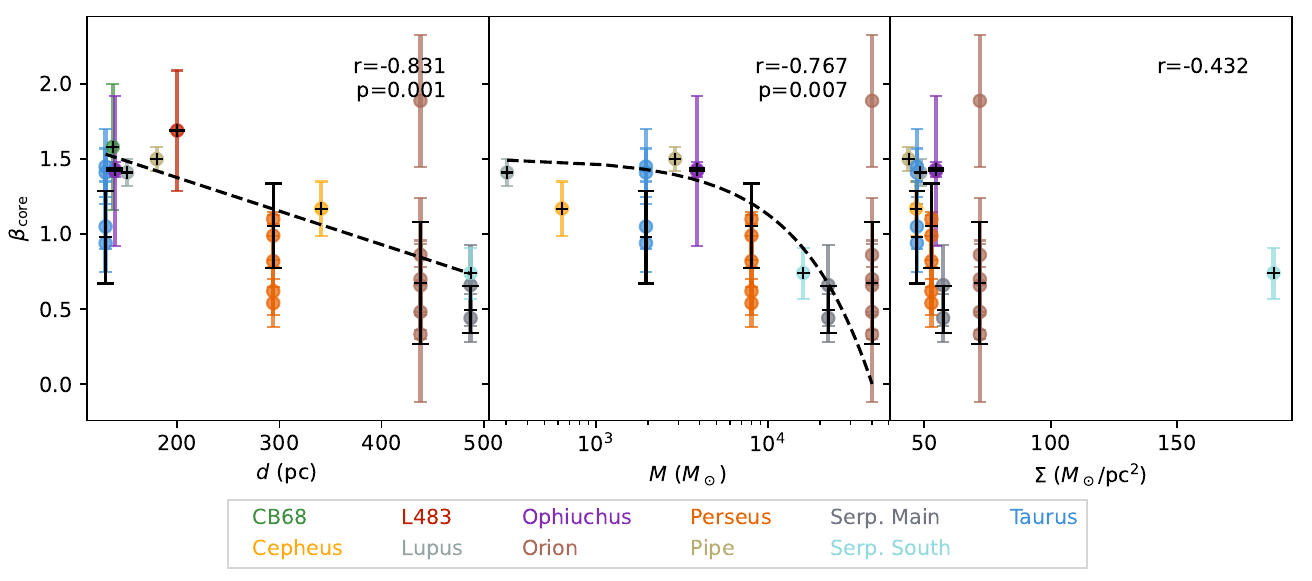}
  \caption[]{Emissivity index of cores observed in \citet{galametz_low_2019,cacciapuoti_faust_2025,nozari_peculiar_2025,bracco_probing_2017}, clustered by region (colour) and plotted relative to distance to the observer (left panel), cloud mass (centre panel), and gas mass surface density (right panel). On each plot, single sources are represented by coloured points. For each region, a weighted average and standard deviation are computed and plotted in black. Measurements in Taurus and Orion from \citet{bracco_probing_2017} and \citet{nozari_peculiar_2025} have been added to examine the most complete observational dataset available (see text for further details). The dashed black line corresponds to inverse-variance \footnotemark linear regression of the black points. P-value $p$ is computed with unweighted linear regression against values uniformly distributed in the range of our data. Correlation coefficient $r$ is also given.}
  \label{fig:Beta_obs_region}
\end{figure*}
\footnotetext{For homogeneity of the data, in this case, the deviation is taken as the average of the deviation in the region.}

To further investigate the possible physical causes for the spread in observed emissivity indices from observations of protostellar cores (grey curves in \cref{fig:Beta_modVSobs}), we sorted each protostellar observation according to the region or molecular cloud it belongs to. We supplemented observations from \citet{galametz_low_2019,cacciapuoti_faust_2025} with observations of Orion protostars from \citet{nozari_peculiar_2025}, and observations of Taurus protostars from \citet{bracco_probing_2017}. Details on harmonisation of these data sources are given in \cref{sec:clouds}. The data is shown in \cref{fig:Beta_obs_region}.
Note that several regions in this plot are sampled by a single source, and thus the statistical significance is limited. However, it seems that the values of the emissivity index are more clustered within a single region than the dispersion found when intercomparing all star-forming regions. The lowest values of $\beta$ on average (\qty{0.5(0.16)}{}), for example, are reached in Serpens Main sources, while the region with the highest $\beta$ (among those with at least two sources observed) is Ophiuchus with a value of \qty{1.43(0.01)}{}.
More specifically, we can use known properties of the regions to which the sources in the sample belong to investigate whether a correlation can be found between the observed emissivity index and the cloud properties (see properties in \cref{tab:cloudprop}). We note a significant correlation ($r>0.5$) of the dust emissivity index with both the distance of the cloud (from the solar system) and, more marginally, with the gas mass of the cloud (see the centre and left panels of \cref{fig:Beta_obs_region}). 
This intriguing gradient of the emissivity index, which decreases with larger cloud distances $d$, is difficult to interpret from a physical point of view, unless the cloud mass is driving the correlation and propagating in the distance correlation as well, since the most massive clouds in our sample are located at larger distances.
We note that, interestingly, the sources from Orion and Serpens South seem to be the major drivers of the correlation in the lower $\beta$ values. It is important to note that these sources are in regions of a high UV field $G_0$ \citep{gutermuth_spitzer_2008,xia_distribution_2022}, which can influence, for example,  dust temperature, and this matter should be investigated further.
We cannot, at this point, provide a definite explanation for this correlation, and specifically we do not rule out an observational bias. We stress that further study should be conducted with an effort to populate this diagram with more observational points to investigate if this tentative correlation holds and explore its origins if it does. Larger samples of the dust emissivity in protostars should be observed to confirm or discard this tentative correlation.

Emissivity index differences from one region to another could be explained by a different dust reservoir at large scales, and inhomogeneities in dust composition and associated materials. From a general point of view, variations in metallicity \citep{de_cia_large_2021}, abundance of silicate and carbonaceous materials \citep{parvathi_probing_2012,mishra_interstellar_2017,zeegers_investigating_2025,decleir_first_2025}, and hence in the composition of dust in the \ac{ism} are documented. \citet{yasuda_co_2023} points out that nearby galaxies exhibit a significant variation with galactocentric radius of their CO-to-H$_2$ abundance ratio and dust-to-gas ratio. This is also observed in our own Galaxy, and this increase in CO abundance towards the inner regions of galaxies is often interpreted as stemming from the radial metallicity gradient from the Galactic centre to the outer Galaxy \citep{kohno_co--h2_2023}.
Indeed, metallicity gradients are observed in our Galaxy \citep{de_cia_large_2021}, suggesting that mixing phenomena in between dust clouds may be largely inefficient, although some recent studies also suggest that the metallicity in the solar neighbourhood may be relatively homogeneous \citep{ritchey_distribution_2023}.
In our sample, the protostars considered are all close to the Sun (distances <\qty{500}{pc}) and thus sample a narrow range of galactocentric radii ($\sim \qty{8\pm1}{kpc}$): it is therefore unlikely that radial gradients with galactocentric distance cause the observed variations.

Typical mixing timescales in the galaxy are usually estimated to be of the order of the galactic orbital period ($\sim 180$ Myr at 8 kpc, \citealt{edmunds_is_1975}), which would suggest a largely inefficient process of mixing compared to the typical cloud lifetimes (10-30 Myr, \citealt{chevance_lifecycle_2019}). However, some more efficient small-scale processes may be at work to provide mixing of galactic materials, bringing the typical mixing timescales to $\sim 15-20$ Myrs (see, for example \citealt{petit_mixing_2015}). Since these timescales are close to the typical crossing times in molecular clouds, yet longer than the star-formation timescales to transform dense gas in young pre-main-sequence stars ($\sim$ 2 Myr, \citealt{evans_spitzer_2009}), it is possible that inhomogeneities persistent at cloud scales could explain the cloud-to-cloud differences, while smaller dispersions of the dust emissivity is observed among protostars quite closely located in the same host cloud. 
We stress that if dust emissivity index differences observed are due to chemical inhomogeneities persisting in the Milky Way, the correlation found with the distance and mass of the star-forming cloud, highlighted in \cref{fig:Beta_obs_region}, remains unexplained, and one must not rule out an observational bias of unknown nature.

\section{Conclusions}
\label{sec:Conclusions}

We studied the emissivity index observed at millimetre wavelength in protostellar environments, confronting predictions from radiative transfer computation of the dust thermal emission and observations. We focused on the variation of the emissivity index from one observed source to another and on expected differences considering one set of optical properties or the other. The main results are described below.

   \begin{enumerate}
      \item We tested different dust optical models, which reflect variations in dust composition, in radiative transfer computations. In each case, we proved capable of retrieving the correct input dust emissivity index from synthetic millimetre observations in the NOEMA and ALMA bands. Importantly, our work allows us to validate the observational method within a 0.1 error range, suggesting that the observed diversity of millimetre dust $\beta$ reported in protostars in the literature must stem from a diversity in intrinsic dust emissivities.\\
      \item When comparing observations of protostars, differences in the emissivity index from one object to another could possibly be explained by dust reservoirs of varying composition. This is supported by a tentative dependence of the emissivity index on the cloud hosting the protostars observed, even if potential observational bias would require further investigation and is yet to be ruled out. Besides, lack of efficient mixing between regions can be a source of inhomogeneous material.\\
      \item The radiative transfer computations we carried out with dust models for diffuse medium are not providing emissivity indices as low as those found in some protostar observations. Dust evolution (e.g. size, aggregation) could still be required to explain such low indices in some objects. An important consideration for future models will be to build models on recent optical properties calibrated against the latest observations of the diffuse medium. Models of dust size evolution in the dense medium could provide a significant contribution in this regard. \\
      \item Within the same dust reservoir, for example, in the observation of a single source, where dust composition is not expected to vary, variations of the emissivity index can be used to trace dust evolution. However, when comparing two different protostars, potentially coming from different dust reservoirs with different atomic abundances, one must be cautious when interpreting low values of observed dust emissivity as resulting from a change in dust grain sizes.
   \end{enumerate}

This work shows that dust properties can be measured with multi-wavelength observations of the thermal dust emission, even in the dense and complex environment of embedded protostars. 
We stress that some of the variety of currently observed emissivity indices in nearby protostars could be intrinsically due to a variety of dust compositions in different star-forming regions. Yet some of the very low values observed can still not be explained using current dust models without dust evolution. In the scope of interpretation of observations, models for dust optical properties are crucial and should be cautiously defined. A complete model, accurate for representing the diversity and early evolution of dust properties in a dense medium, is lacking and will be the topic of future work.

\begin{acknowledgements}
This work was made possible thanks to the support from the European Research Council (ERC) under the European Union’s Horizon 2020 research and innovation programme (Grant agreement No. 101098309 - PEBBLES).
Colour sequences in plots of this project are designed to be accessible thanks to \citet{petroff_accessible_2021}.
This research has made use of data from the Herschel Gould Belt survey (HGBS) project (http://gouldbelt-herschel.cea.fr). The HGBS is a Herschel Key Programme jointly carried out by SPIRE Specialist Astronomy Group 3 (SAG 3), scientists of several institutes in the PACS Consortium (CEA Saclay, INAF-IFSI Rome and INAF-Arcetri, KU Leuven, MPIA Heidelberg), and scientists of the Herschel Science Center (HSC). A.M. acknowledges support by the program Unidad de Excelencia María de
Maeztu, awarded to the Institut de Ciències de l’Espai (CEX2020-001058-M).
\end{acknowledgements}

\bibliographystyle{aa} 
\bibliography{bibliography}

\begin{appendix}

\section{Template MHD model of protostellar core}
\label{sec:MHDdescr}

The \ac{mhd} model used in this study is taken from simulations from \cite{hennebelle_what_2020}, which use the \ac{amr} code RAMSES \citep{teyssier_cosmological_2002} to compute the collapse of a 1 $M_\odot$ molecular core. The complete description of the simulation can be found in \cite{valdivia_is_2022} (our model corresponds to the line "ID R1" in this reference). From this simulation, we select the output at which the sink particle has reached a total mass of $0.1M_\odot$, 2493 yrs after its formation. We study this simulation from an ``edge-on'' view, considering the line of sight along the $y$-axis. We plot the dust density profile in \cref{fig:Dens}, and we can note the \qty{30}{\degree} angle between the initial magnetic field ($z$ axis) and the rotation axis of the system.

We use this model as a template: we stress that the whole radiative transfer modelling does not aim to represent a realistic protostar. The dust models used in this work, by their size and optical properties, are representative of the diffuse \ac{ism} rather than a protostellar envelope. We study here solely the influence that dust optical properties can have on the observables, hence the precise characteristics of the \ac{mhd} model have little impact.

\section{Dust size distribution function}
\label{sec:dustsize}

In this section, we discuss the choice of dust size distribution used in this study. We stress that the choice of grain size distribution in a model should be motivated by physical processes responsible for such a grain population. Distribution of dust by power-law \citep{mathis_size_1977} reproduces well the resulting population from collisional fragmentation \citep{dohnanyi_collisional_1969,tanaka_steady-state_1996}, whereas grown grains tend to deviate from this behaviour \citep{birnstiel_dust_2011}, and collisional growth tends to produce log-normal distributed populations of grains \citep{ysard_grains_2019, lorek_local_2018}. As explained in \cref{subsec:dust}, we populated our radiative transfer computation with a log-normal distribution of dust grains, as we believe it is likely that, in these dense environments, grains already underwent some collisional growth. Here we compare this hypothesis with the commonly used MRN \citep{mathis_size_1977} distribution, namely a $-3.5$ power-law size distribution in number of grains.

Dust distribution in mass, number and extinction at \qty{1}{\mm} of the log-normal and MRN models are respectively plotted \cref{fig:Size_dis_logn} and \cref{fig:Size_dis_MRN}. The two distributions show a significant difference in the repartition of the number of grains and mass repartition throughout the size range. However, those differences have a negligible impact on the resulting emissivity index. We compute, as detailed more precisely in \cref{sec:Results}, the emissivity index from intensity $I$ maps as : $$\beta = \frac{\log I_{\nu_1}-\log I_{\nu_2}}{\log \nu_1 - \log \nu_2} - \frac{\log B_{\nu_1}-\log B_{\nu_2}}{\log \nu_1 - \log \nu_2}$$ with $B$ the estimated Planck function. \cref{fig:Beta_MRNvslogn} compares the (zoomed in) radial emissivity index resulting from the radiative transfer of THEMIS 2.0 grains with both size distributions. We note that the difference of the emissivity index between the two distributions remains always smaller than 1\%.  The choice of dust distribution thus doesn't change the results of this study and does not impact the variability of emissivity indices we find when using different dust models.

\begin{figure}[t]
  \centering
  \includegraphics[width=\linewidth,clip]{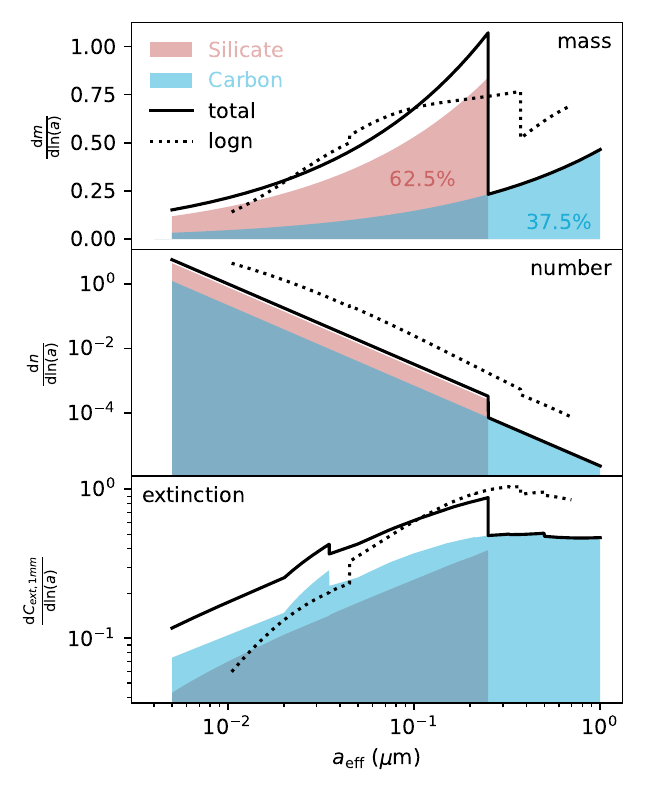}
  \caption{Dust distributions in mass, number and $C_{\mathrm{ext}}$ at \qty{1}{\mm}. The distributions are normalised, such that the integral of the total distribution equals 1 in each panel. The distribution stems from a power law (MRN) in number. In dotted line, we plot the total distribution for the log-n case (see \cref{fig:Size_dis_logn}).}
  \label{fig:Size_dis_MRN}
\end{figure}

\begin{figure}[t]
  \centering
  \includegraphics[width=\linewidth,clip]{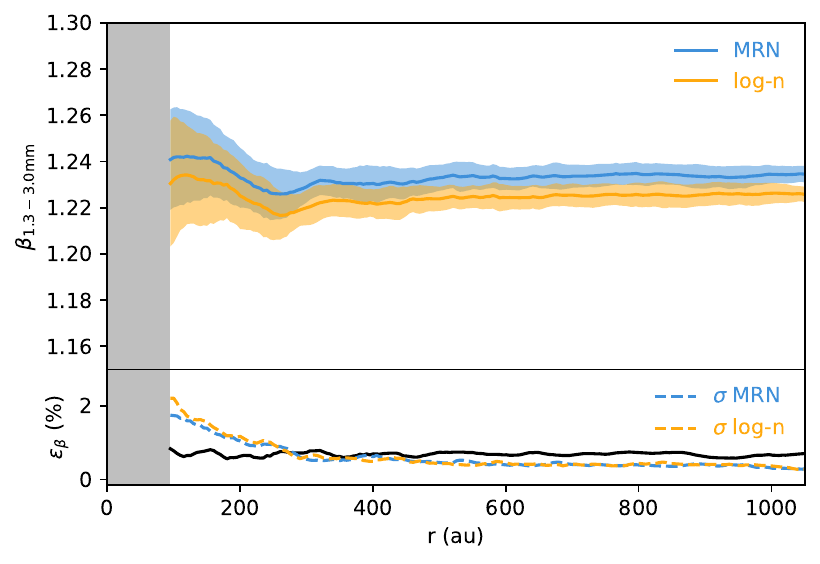}
  \caption{Comparison of the radial emissivity index for the THEMIS 2.0 model between two dust size distributions. Bottom panel shows relative error $\epsilon_\beta$ between the two curves of the top panel.}
  \label{fig:Beta_MRNvslogn}
\end{figure}

\section{Stoichiometry effects}
\label{sec:asil}

\begin{figure}[t]
  \centering
  \includegraphics[width=\linewidth,clip]{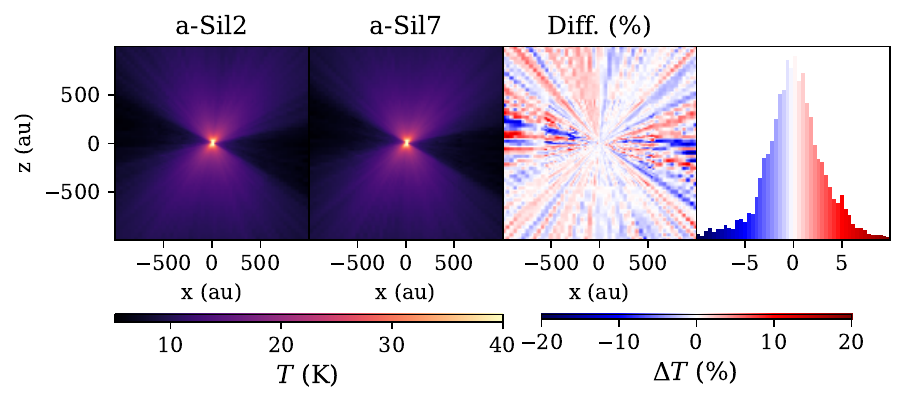}
  \caption{Comparison of the temperatures resulting from the radiative transfer computation, including THEMIS 2.0 dust model with a-Sil2 and a-Sil7 material, respectively. The second right panel shows the relative difference between the left panel maps, and the far right panel plots the histogram of the differences.}
  \label{fig:aSilT}
\end{figure}

\begin{figure}[t]
  \centering
  \includegraphics[width=\linewidth,clip]{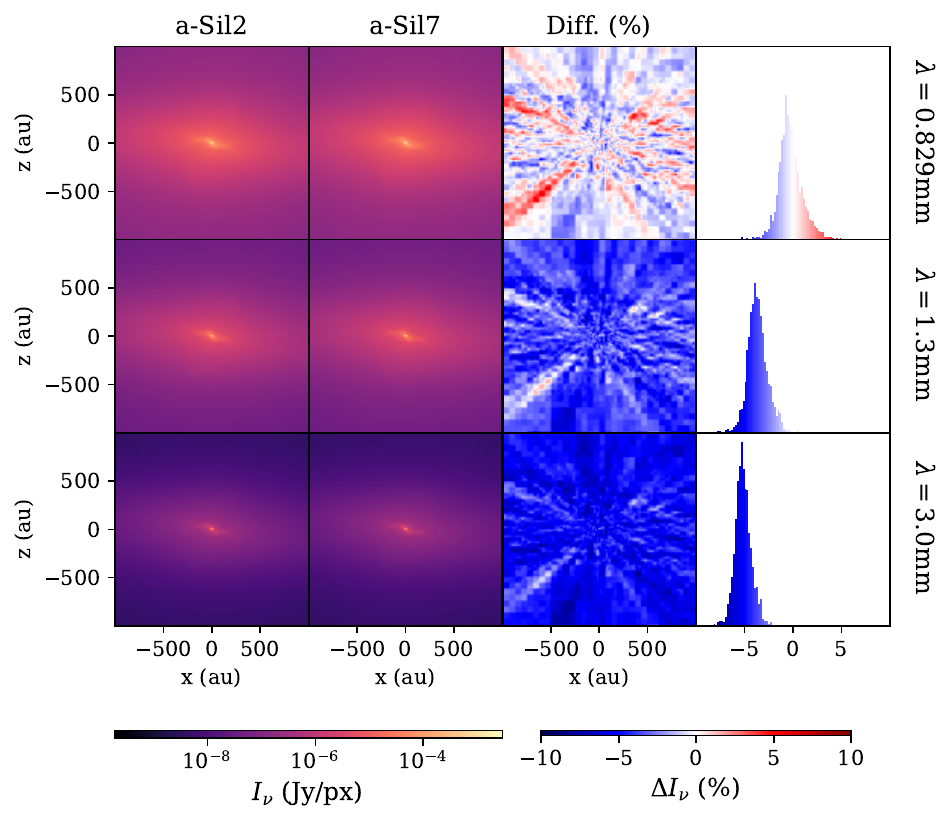}
  \caption{Comparison of the flux density maps resulting from the radiative transfer computation, including THEMIS 2.0 dust model with a-Sil2 and a-Sil7 material, respectively. The second right panel shows the relative difference between the left panel maps, and the far right panel plots the histogram of the differences.}
  \label{fig:aSilI}
  
\end{figure}
\begin{figure}[t]
  \centering
  \includegraphics[width=\linewidth,clip]{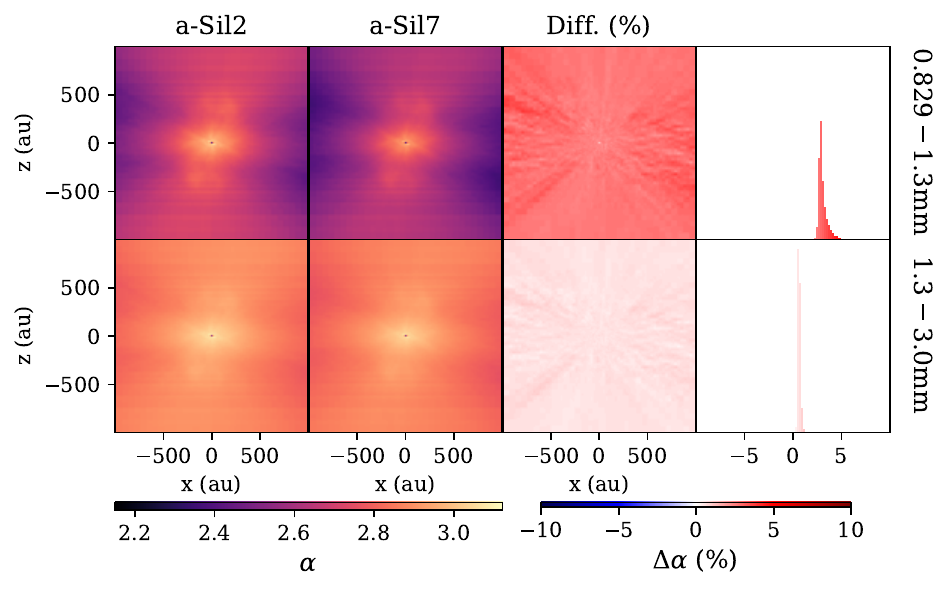}
  \caption{Comparison of the spectral index maps computed from the flux density \cref{fig:aSilI}, including THEMIS 2.0 dust model with a-Sil2 and a-Sil7 material, respectively. The second right panel shows the relative difference between the left panel maps, and the far right panel plots the histogram of the differences.}
  \label{fig:aSilBetas}
\end{figure}

This section goes into even more subtle variations in the dust composition that could induce differences in the dust emissivity index. In \cref{subsec:dust_on_beta}, we discussed the effect of the dust model on the value of the emissivity index retrieved in radiative transfer computations. Rather than changing the whole dust model, here we compare two different stoichiometries for the silicate material in THEMIS 2.0 \citep{ysard_themis_2024}, namely a-Sil2 (from the best fit model, used hereinbefore) and a-Sil7. Note that the materials used here are the two extreme silicates of THEMIS 2.0 in terms of spectral index (see Table 2 in \citealt{ysard_themis_2024}). 
\Cref{fig:aSilT} plots the comparison of temperatures resulting from both radiative transfer computations. There is no significant deviation of temperature in between the two computations\footnote{The variations of the order of 5\% seen in the last panel of \cref{fig:aSilT} are typical of the noise due to the Monte Carlo dust heating method in our computations.}.
Even if dust heating mechanisms are not influenced by the change of dust properties in this study case, dust emission can be impacted by dust intrinsic properties. Looking at dust emission \cref{fig:aSilI}, we notice characterised differences in intensity at wavelengths larger than \qty{1}{\mm}, with lower emission for the a-Sil2 model, and absolute difference increasing with wavelength. This eventually results in spectral index differences, as one can notice in \cref{fig:aSilBetas}. As predicted by characteristics of silicate grains detailed in \citet{ysard_themis_2024}, the population including a-Sil7 material produces lower emissivity indices than computations with a-Sil2 material. 
This comparison proves the importance of the smallest variation of the composition of dust grains and the influence it can have on the observed emissivity index. It is, though, important to note that the influence of the stoichiometry of the silicate material remains small compared to the influence of the whole dust model composition as seen in \cref{subsec:dust_on_beta}.

\section{Effects of temperature considerations on the emissivity index}
\label{sec:temperature}

\begin{figure}[t]
  \centering
  \includegraphics[width=\linewidth,clip]{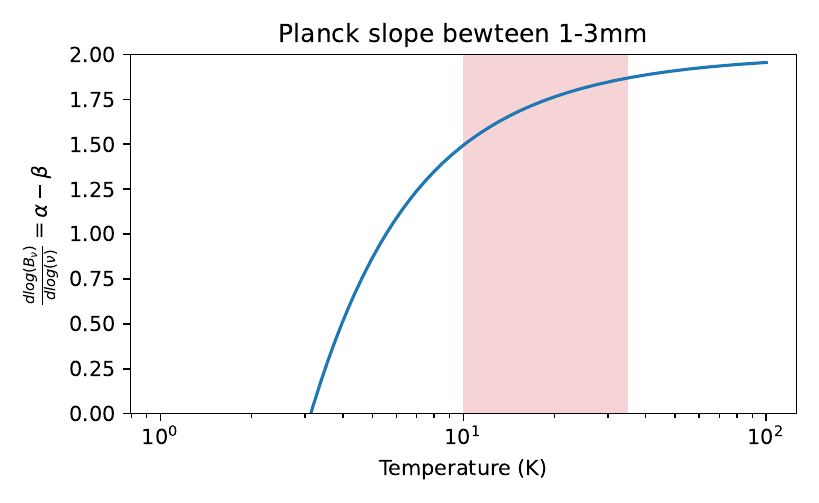}
  \caption{Slope of the Planck function between \qty{99.93}{\GHz} (\qty{3}{mm}) and \qty{230.61}{\GHz} (\qty{1}{\mm}) depending on the temperature. In red is the temperature range in the computation considered here.}
  \label{fig:Planck}
\end{figure}

\begin{figure*}[t]
  \centering
  \includegraphics[width=\linewidth,clip]{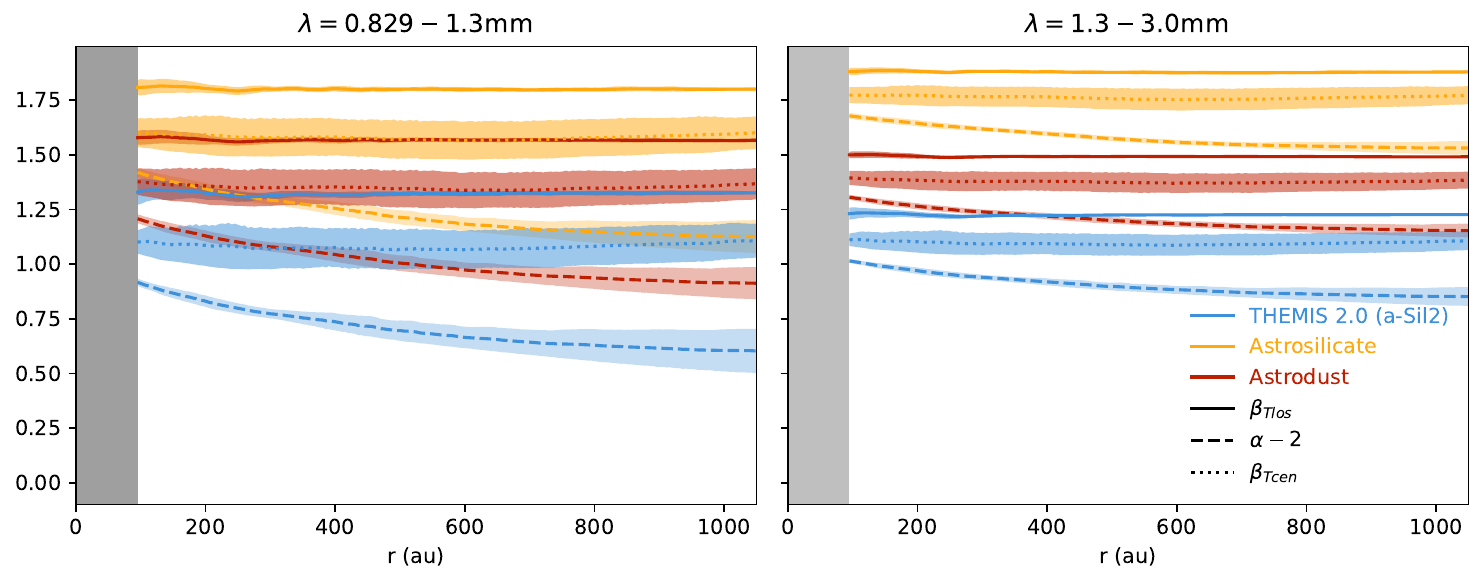}
  \caption{Comparison of the radial profile of the dust emissivity index $\beta$ computed from the spectral index as $\alpha-2$ (dashed lines, Rayleigh-Jeans approximation), the dust emissivity index $\beta$ computed with the Planck function averaged on the line of sight (solid lines), and the dust emissivity index $\beta$ computed with the Planck function using the temperature on the centre plane of the simulation (dotted lines).}
  \label{fig:BetavsAlpha}
\end{figure*}

When computing the emissivity index, we mention in \cref{sec:Results} the invalidity of the Rayleigh-Jeans approximation in our case. Indeed, the Planck function can be approximated as a square function of the frequency at long wavelengths, but the colder the temperature, the further we stand from this power-law at a certain wavelength. For instance, \cref{fig:Planck} shows the slope of the Planck function between \qty{99.93}{\GHz} (\qty{3}{mm}) and \qty{230.61}{\GHz} (\qty{1}{\mm})\footnote{Wavelengths we use for the $\beta$ calculation in \cref{fig:Beta_modVSobs}.} relatively to the temperature. In this case, we notice that the Rayleigh-Jeans approximation is only valid when approaching \qty{100}{\K}, and the slope is largely deviating from 2 at colder temperatures. In our temperature ranges, the Planck slope ranges from 1.85 to 1.5. 
Consequently, using the Rayleigh-Jeans approximation $\beta_\mathrm{R-J} = \alpha - 2$ could lead to two major errors in the emissivity index:
\begin{itemize}
    \item under-estimation of $\beta$ because of general low temperatures;
    \item incorrect spatial variations of $\beta$ due to temperature fluctuations. 
\end{itemize}
We illustrate those discrepancies in \cref{fig:BetavsAlpha}, which plots both the emissivity index $\beta$ computed as described in \cref{eq:beta}, and the index $\beta_\mathrm{R-J}=\alpha-2$ computed in the Rayleigh-Jeans approximation. We do see the overall under-estimation of the index when using the Rayleigh-Jeans approximation, and a spatial structure -- namely, a negative gradient with the radius -- resulting from the radial dependence of the temperature. Retrieving the temperature effects along the line of sight in a model is quite straightforward since all parameters of the 3D grid are accessible. However, in observations, accessing the temperature can be a challenge and particular care is required when correcting the temperature effects on the emissivity index.

\section{Convolved maps}
\label{sec:conv}

In this appendix, we provide maps of flux density (Fig. \ref{fig:IMapsBeam}) and spectral index $\alpha$ (Fig. \ref{fig:AlphaBeam}) convolved by a 0.5\arcsec \,Gaussian beam. Those maps thus provide the flux densities from thermal dust emission, with units comparable to the ones in the maps obtained from, e.g., interferometric observations.

\begin{figure}[t]
  \centering
  \includegraphics[width=\linewidth,clip]{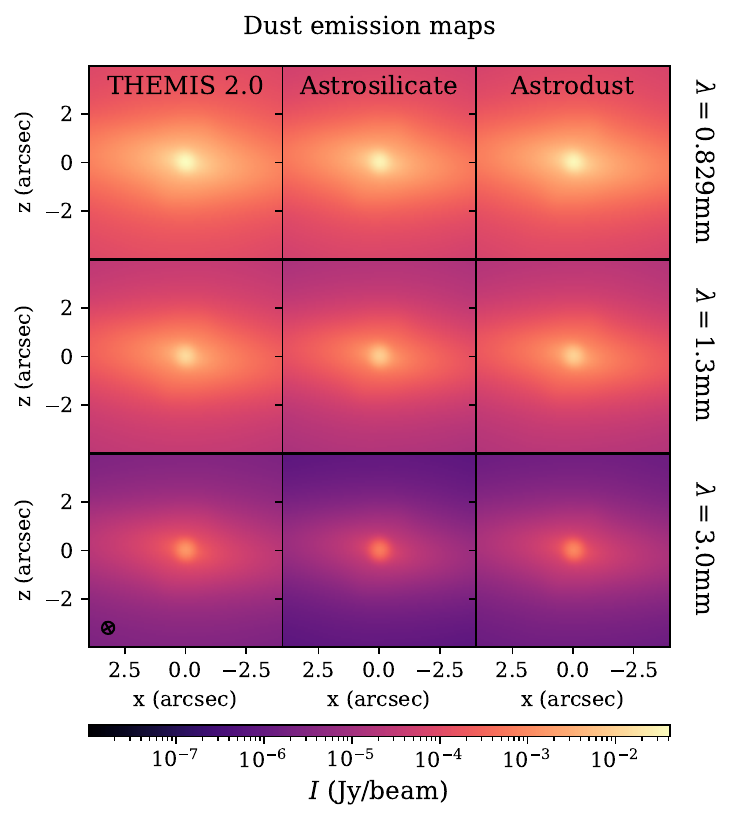}
  \caption{Flux density maps for the dust thermal emission at the three wavelengths considered in our study (from top to bottom \qty{0.8}{\mm} - ALMA Band7, \qty{1.3}{\mm} - ALMA Band6, and \qty{3}{\mm} - ALMA Band3). Each map is convolved with a 0.5 arcsecond Gaussian beam. In each column, a different dust model is used for the radiative transfer computation (from left to right THEMIS 2.0, Astrosilicate and Astrodust).}
  \label{fig:IMapsBeam}
\end{figure}

\begin{figure}[t]
  \centering
  \includegraphics[width=\linewidth,clip]{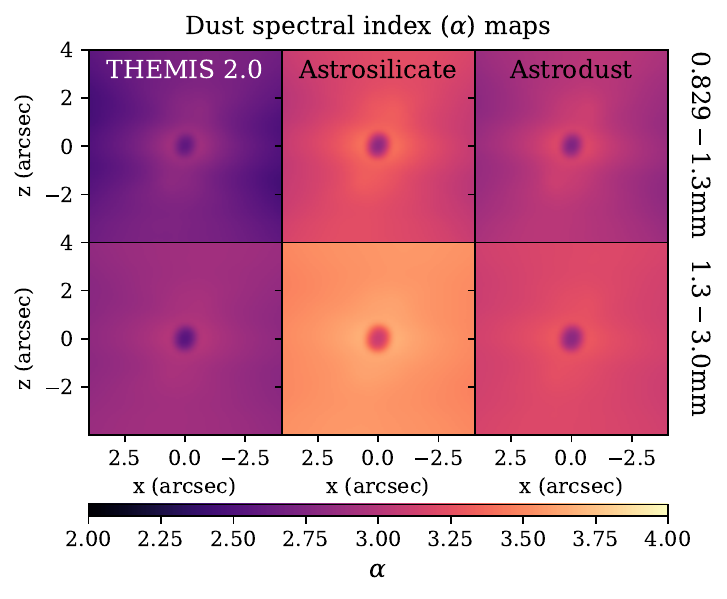}
  \caption{Spectral index $\alpha$ maps, computed directly from the ratio of the convolved flux density maps (shown in \cref{fig:IMapsBeam}), over different wavelength ranges (from top to bottom \qty{0.8}{\mm}-\qty{1.3}{\mm} and \qty{1.3}{\mm}-\qty{3}{\mm}). In each column, a different dust model is used (from left to right THEMIS 2.0, Astrosilicate and Astrodust).}
  \label{fig:AlphaBeam}
\end{figure}

\section{Investigation on cloud properties - supplementary material}
\label{sec:clouds}

This appendix provides supplementary material on the discussion raised in \cref{subsec:inhomogeneity}, detailed in \cref{tab:cloudprop,fig:cloud_prop}.

As shown by our modelling work, the Rayleigh-Jeans approximation $\beta = \alpha - 2$ is inaccurate in the inner envelope of protostars because of temperature effects: therefore, measurements from \citet{nozari_peculiar_2025} in Orion protostars used here and in \cref{subsec:inhomogeneity} are the $\alpha-\mathrm{ALMA}$ values (see their Table 4), from which we recompute the $\beta$ according to the formula given in \citet{galametz_low_2019} (see their Fig.\,4). These $\beta$ values are subsequently corrected to account for the optically-thick contribution from the central region (subtractive factor of $-0.13$ as the mean correction found for Serpens protostars in \citealt{galametz_low_2019}). 
We note that among those Orion sources, one high $\beta$ value of 1.88 stands out. This value corresponds to FIR2, which displays an atypical SED (see Fig. 3 of \citealt{nozari_peculiar_2025}), which could suggest an important non-thermal component, contrary to the other sources of this study (as stated by the authors themselves). The SED of this source implies either a significant change in emissivity between 2 and 3~mm, or that the measure of $\beta$ in this source is inconclusive due to its complexity.
Measurements from \citet{bracco_probing_2017} for two Taurus protostars in B213 are also considered, using the $\beta$ values they report at the IRAM-30m beam size (1700 au at the Taurus distance, see their Fig. 5).

We note that we find no evident correlation (correlation coefficient $r<0.5$) between the emissivity index of the dust in the protostars and the gas mass surface density of the cloud, nor with the star formation efficiency or fraction of dense gas of the cloud (see right panel of \cref{fig:Beta_obs_region} and \cref{fig:cloud_prop}).

\begin{table*}[ht]
\centering
\caption{Main properties of the regions (or source for isolated protostars) where the observed sources are located.}
\label{tab:cloudprop}
\begin{tabular}{lccccccc}
\hline
Cloud                              & $d$\tablefootmark{(a)} & $M$\tablefootmark{(a)}   & $L_{\rm Av1}$\tablefootmark{(b)} & $\Sigma$                      & $f_{\rm dg}^N$\tablefootmark{(a)} & $f_{\rm dg}^\rho$\tablefootmark{(a)} & SFE\tablefootmark{(a)} \\
                                   & [pc]                 & [$M_\odot$]            & [pc$^2$]                       & [$M_\odot\,\mathrm{pc}^{-2}$] & [\%]                            & [\%]                               & [\%]                 \\ \hline
Cep1157                            & 341                  & 633                    & 13.53                          & 46.78                         & 4.85                            & 0.66                               & 0.47                 \\
Lup I                              & 151                  & 304                    & 6.27                           & 48.48                         & 5.13                            & 1.60                               & 1.78                 \\
Oph L1688                          & 139                  & 3853                   & 70.56                          & 54.61                         & 12                              & 2.25                               & 3.0                  \\
Orion A                            & 438                  & 40072                  & 556.83                         & 71.96                         & 24.9                            & 3.4                                & 3.19                 \\
Perseus                            & 294                  & 7990                   & 151.25                         & 52.83                         & 13.36                           & 2.9                                & 2.11                 \\
Pipe Nebula                        & 180                  & 2881                   & 65.79                          & 43.79                         & 0.81                            & 0.29                               & 0.29                 \\
Serp. Main                         & 487                  & 22307\tablefootmark{(b)} & 388.14                         & 57.47                         & ---                             & ---                                & ---                  \\
Serp. South                        & 487                  & 15964\tablefootmark{(b)} & 84.86                          & 188.10                        & ---                             & ---                                & ---                  \\
Tau L1495                          & 130                  & 1948                   & 41.39                          & 47.06                         & 5.6                             & 0.88                               & 2.11                 \\ \hline
Source                             & $d$\tablefootmark{(c)}                  & \multicolumn{1}{l}{}   & \multicolumn{1}{l}{}           & \multicolumn{1}{l}{}          & \multicolumn{1}{l}{}            & \multicolumn{1}{l}{}               & \multicolumn{1}{l}{} \\
                                   & [pc]                 & \multicolumn{1}{l}{}   & \multicolumn{1}{l}{}           & \multicolumn{1}{l}{}          & \multicolumn{1}{l}{}            & \multicolumn{1}{l}{}               & \multicolumn{1}{l}{} \\ \hline
CB68                               & 137                  & \multicolumn{1}{l}{}   & \multicolumn{1}{l}{}           & \multicolumn{1}{l}{}          & \multicolumn{1}{l}{}            & \multicolumn{1}{l}{}               & \multicolumn{1}{l}{} \\
L483                               & 200                  & \multicolumn{1}{l}{}   & \multicolumn{1}{l}{}           & \multicolumn{1}{l}{}          & \multicolumn{1}{l}{}            & \multicolumn{1}{l}{}               & \multicolumn{1}{l}{} \\ \hline
\end{tabular}
\tablefoot{\tablefoottext{a}{From \citet{orkisz_volume_2025}}
\tablefoottext{b}{Computed by integration of the Herschel Gould Belt Survey maps \citep{andre_filamentary_2010} considering $A_V$>1 regions.}
\tablefoottext{c}{\citet{cacciapuoti_faust_2025}}}
\end{table*}

\begin{figure*}[h]
\centering
\begin{subfigure}{.33\textwidth}
  \centering
  \includegraphics[width=\linewidth]{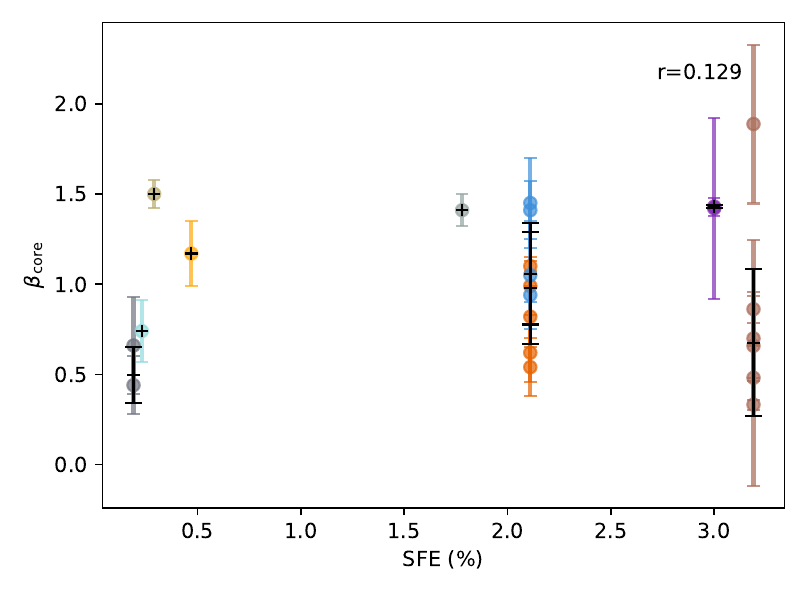}
\end{subfigure}
\begin{subfigure}{.33\textwidth}
  \centering
  \includegraphics[width=\linewidth]{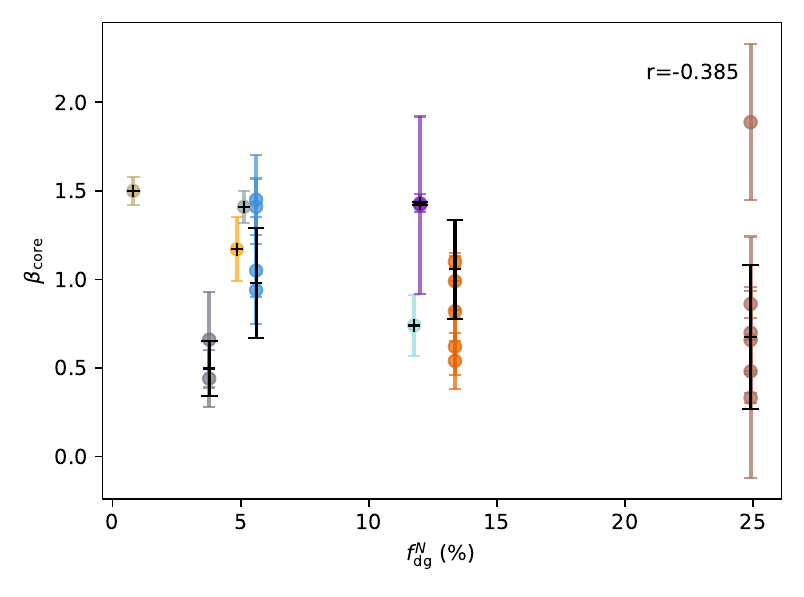}
\end{subfigure}
\begin{subfigure}{.33\textwidth}
  \centering
  \includegraphics[width=\linewidth]{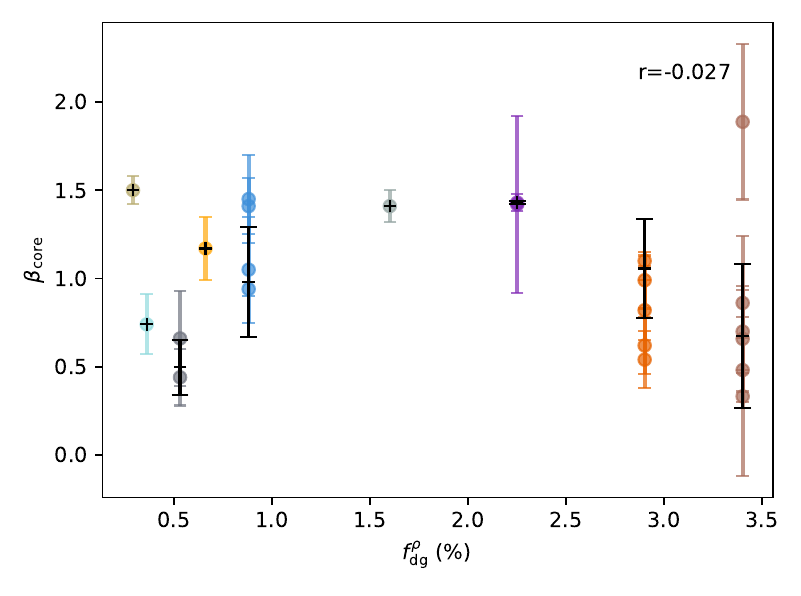}
\end{subfigure}
\caption{Emissivity index in protostellar cores observed in \citet{galametz_low_2019,cacciapuoti_faust_2025,nozari_peculiar_2025,bracco_probing_2017}, clustered by region (colour) and plotted relative to star formation efficiency (SFE) (top left panel), column-density-based dense gas fraction (top right panel) and volume-density-based dense gas fraction (bottom panel). On each plot, single sources are represented by coloured points. Measurements in Taurus and Orion from \citet{bracco_probing_2017} and \citet{nozari_peculiar_2025} have been added to examine the most complete observational dataset available, see text for further details. For each region, a weighted average and standard deviation are computed and plotted in black.  Correlation coefficient $r$ is given.}
\label{fig:cloud_prop}
\end{figure*}

\end{appendix}

\end{document}